\documentclass{iopart}
\usepackage{iopams}
\usepackage{graphicx}
\usepackage{units}
\usepackage{ulem}
\usepackage{color}
\usepackage[latin1]{inputenc}
\usepackage{cite}
\expandafter\let\csname equation*\endcsname\relax
\expandafter\let\csname endequation*\endcsname\relax 
\usepackage{amsmath}
\usepackage{booktabs}

\begin{document}

\title{Mapping the Optical Absorption of a Substrate-Transferred Crystalline AlGaAs Coating at \unit[1.5]{$\mu$m}}

\author{Jessica Steinlechner$^1$, Iain W Martin$^1$, Angus Bell$^1$, Garrett Cole$^2$, Jim Hough$^1$, Steven Penn$^3$, Sheila Rowan$^1$, Sebastian Steinlechner$^1$}

\address{$^1$ SUPA, School of Physics and Astronomy, University of Glasgow, Glasgow,
G12 8QQ, Scotland}

\address{$^2$ Faculty of Physics, University of Vienna, Boltzmanngasse 5, 1090 Vienna, Austria}

\address{$^3$ Department of Physics, Hobart and William Smith Colleges, Geneva, NY 14456, USA}

\ead{iain.martin@glasgow.ac.uk}

\date{\today}
%
%
\begin{abstract}

The sensitivity of 2nd and 3rd generations of interferometric gravitational wave detectors will be limited by thermal noise of the test-mass mirrors and highly reflective coatings. Recently developed crystalline coatings show a promising thermal noise reduction compared to presently used amorphous coatings. However, stringent requirements apply to the optical properties of the coatings as well. We have mapped the optical absorption of a crystalline AlGaAs coating which is optimized for high reflectivity for a wavelength of \unit[1064]{nm}. The absorption was measured at \unit[1530]{nm} where the coating stack transmits approximately \unit[70]{\%} of the laser light. The measured absorption was lower than $\unit[(30.2 \pm 11.1)]{ppm}$ which is equivalent to $\unit[(3.6 \pm 1.3)]{ppm}$ for a coating stack that is highly reflective at \unit[1530]{nm}. While this is a very promising low absorption result for alternative low--loss coating materials, further work will be necessary to reach the requirements of $< \unit[1]{ppm}$ for future gravitational wave detectors.

\end{abstract}

%

\section{Introduction}

Gravitational wave (GW) detectors are the most sensitive displacement measurement devices in the history of physics. Large Michelson interferometers with perpendicular arms were developed to search for extremely small changes in the relative distance of test-mass mirrors. These mirrors are coated to be highly reflective and form arm cavities of several kilometers in length~\cite{Pitkin2011}. The 2nd generation of these detectors with improved sensitivity~\cite{aligo, avirgo} is presently being implemented and is expected to make the first detection of GWs within the next two years. In the most sensitive part of the detection bandwidth, the performance of these detectors is expected to be limited by thermal noise from the optical coatings applied to the test-mass mirrors.

To reduce the impact of thermal noise on the system performance, future detectors like \textit{Advanced LIGO} (aLIGO) upgrades and the \textit{Einstein Telescope} (ET) are planned to operate at cryogenic temperatures~\cite{Hild2010, Adhikari2012}. The requirements on the mirror-substrate material and the coatings are low mechanical loss to minimize Brownian thermal noise, low optical absorption to maintain the cryogenic base temperature and reduce background heating, as well as to avoid thermal lensing, and very accurately specified high reflectivity~\cite{Harry2007}.
The mechanical loss of the presently used substrate material fused silica (FS) increases towards low temperatures~\cite{Fine1954} making FS unsuitable for cryogenic GW detector application. Crystalline silicon (c-Si) has been identified as a promising replacement material, with several attractive properties~\cite{Rowan2003, Winkler1991} including a low intrinsic mechanical loss which decreases at low temperatures~\cite{Nawrodt2008, McGuigan1978}. Because of high optical absorption in c-Si at \unit[1064]{nm} a change in laser wavelength is required towards longer wavelengths where the absorption decreases rapidly~\cite{Keevers1995}. Operation at the telecommunication wavelength of \unit[1550]{nm} has been proposed as the most reasonable approach because of available high quality optical components and stable high power lasers. Amorphous SiO$_2$/Ta$_2$O$_5$ coatings match the absorption requirements at both \unit[1064]{nm} and \unit[1550]{nm} but the mechanical loss is still above the proposed requirements~\cite{Granata2013}.

Recently, crystalline coatings composed of alternating layers of gallium arsenide (GaAs) and aluminium gallium arsenide (AlGaAs) have been developed which show promising thermal noise and absorption properties. At \unit[1064]{nm} an absorption of \unit[12.5]{ppm} and a reduction of the Brownian thermal noise by a factor of three (corresponding to a tenfold reduction in the mechanical loss angle) compared to SiO$_2$/Ta$_2$O$_5$ coatings has been reported~\cite{Cole2013}. These crystalline coatings have to be grown on bulk crystalline substrates with matching coating and substrate crystal structure and lattice parameters. For GW application the coatings then have to be transferred to a suitable test-mass substrate. The cavity input test masses (ITMs) for aLIGO and future GW detectors have design transmissions of about \unit[1]{\%} while the end test-masses (ETMs) have much lower transmission in the order of a few ppm~\cite{aligo, ETdesign}. For ITM transmission the laser beam will see an absorption sum from all potential absorption sources such as the mirror substrate, the substrate surfaces and the interface between the bonded coating and substrate, while the beam is basically unaffected by other sources than pure coating absorption when being reflected by the ETM.

In this paper we report absorption measurements on a GaAs/AlGaAs coating stack at a wavelength of \unit[1530]{nm}. The coating stack was optimized for a reflectivity of $R=\unit[99.9996]{\%}$ ($1-R=\unit[4]{ppm}$) at \unit[1064]{nm} and was grown on a GaAs crystal before being transferred and bonded to a FS disc using a direct bonding procedure~\cite{Cole2013}. We measured the optical absorption of the AlGaAs coating using \textit{Photothermal Common-path Interferometry} (PCI)~\cite{Alexandrovski2009}. In our experiment we have mapped the absorption of a \unit[4]{mm} x \unit[5.7]{mm} area to investigate the uniformity of the coating. The coating stack which is highly reflective at \unit[1064]{nm} transmits about \unit[70]{\%} of the input laser power at \unit[1530]{nm}. The ability to measure in transmission provides information of the absorption of the coating stack including any possible absorption associated with the bond between the coating and the substrate. Subsequently the measured absorption is scaled to allow comparison with the highly reflective case.

\section{The Experiment}

\begin{figure}
  \centering
  \includegraphics[width=8cm]
    {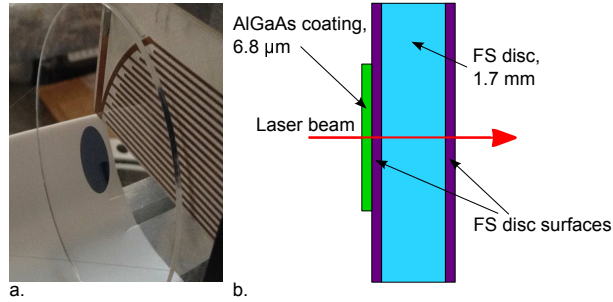}
  \caption{a. Photograph of the \unit[3]{''} diameter fused silica disc with the \unit[16.4]{mm} diameter AlGaAs coating in the middle. b. Schematic of the AlGaAs coated sample: The laser beam transmits through the AlGaAs coating (green), the substrate surfaces (purple), and the bulk of the fused silica substrate (blue).}
  \label{fig:Sample}
\end{figure}

The crystalline AlGaAs coating stack is identical to that described by Cole et al~\cite{Cole2013} for which a direct thermal noise measurement was made. It consists of 40.5 bilayers of alternating GaAs and AlGaAs and is optimized for high reflectivity (HR) at a wavelength of \unit[1064]{nm}. A circular stack \unit[16.4]{mm} in diameter was bonded to a FS disc \unit[3]{''} in diameter and \unit[1.7]{mm} in thickness (see Fig.~\ref{fig:Sample}\,a.). We measured the optical absorption of the coating at a wavelength of \unit[1530]{nm}, where the stack transmits \unit[70]{\%} of the laser power, using the PCI technique~\cite{Alexandrovski2009}.

\subsection{Absorption of the AlGaAs coating}

\begin{figure}
  \centering
  \includegraphics[width=12cm]
    {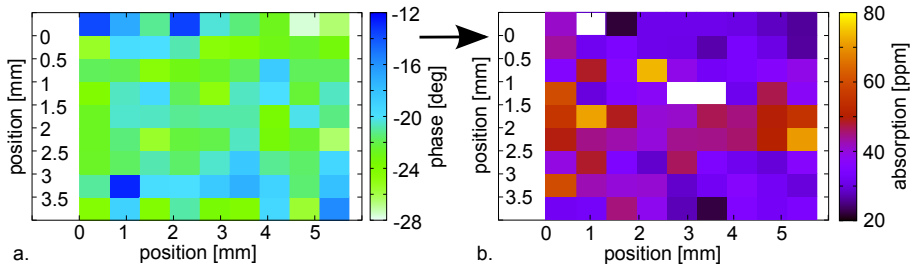}
  \caption{Map of the phase (a.) and absorption (b.) of a \unit[4]{mm} x \unit[5.7]{mm} area on the AlGaAs coating: a. shows a map of the phase as a function of position on the coating. b. shows the optical absorption with an average of $\alpha_{\rm 1530} = \unit[(38.4 \pm 10.3)]{ppm}$ after recalibration considering the phase information. (The white squares indicate high absorbing spots on the coating, presumably originating from dust.)}
  \label{fig:map}
\end{figure}

PCI measures the change in phase of a portion of a large probe beam ($\unit[115]{\mu m}$ radius for the measurements presented here) caused by the absorption of light from a small pump beam ($\unit[35]{\mu m}$). The phase difference created by the pump beam within the probe beam is read out by measuring the interference between the light unaffected by the pump and the light which has been affected by the pump. This interference has a maximum value when measured approximately one (pump) Rayleigh range from the pump waist. The interferometric phase change measured at this point is directly proportional to the absorption of the material.
The pump beam is amplitude modulated at a given frequency and a lock-in amplifier is used to recover the modulation of the interferogram at this frequency. The lock-in also provides the phase of this signal with reference to the pump modulation. This phase is determined by the diffusion of the heat from the pump absorption in the medium and by the pump beam dimensions. In the case where the transverse dimensions are much smaller than the change in intensity along the beam, the characteristic frequency is given by $f = 4 / \pi \times D/(w_0^2)$, where D is the thermal diffusivity of the material and $w_0$ is the beam waist.  For modulation faster than this frequency the phase tends to $\unit[-90]{^\circ}$, whereas for slower frequencies it tends to $\unit[0]{^\circ}$.

For the case of coating absorption, PCI and other methods which are based on heating assume an immediate heat transfer from the coating into the underlying substrate, and that the area of the substrate which is heated and therefore affects the probe beam is large compared to the thin coating~\cite{Willamowski1998, Steinlechner2012}. Amorphous coatings usually show a small thermal expansion coefficient $a_{\rm th}$ and a small thermo-refractive coefficient d$n$/d$T$. So the phase change caused within a thin coating of only a few $\mu$m, is small compared to the phase change caused within the substrate. These assumptions are valid for amorphous, sputtered coatings for which the majority of the signal is caused by the substrate while the signal produced directly in the coating is negligible.

For the AlGaAs coating an area of \unit[4]{mm} x \unit[5.7]{mm} was mapped to check the homogeneity of the absorption. Figure~\ref{fig:map}\,a. shows a map of the phase signal which provides two significant conclusions:

\begin{enumerate}
	\item The phase varies strongly for different positions on the coating.
	\item The mean value of $\unit[-21]{^\circ}$ differs significantly from the expected phase for the FS disc of $\unit[-65]{^\circ}$ for the used chopper frequency of \unit[407]{Hz}.
\end{enumerate}

The variation across the coating indicates a non-uniform heat conduction of coating and substrate (for an assumption of uniform coating and substrate material parameters) which may be related to variations in the quality of the bond between the coating and the substrate. The discrepancy of the mean phase from the expected phase for FS in general shows that the signal from the coating contributes significantly to the total signal.

\begin{figure}
  \centering
  \includegraphics[width=13cm]
    {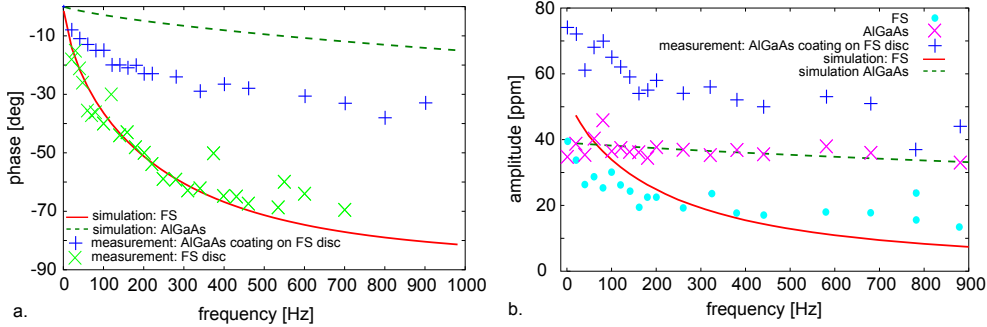}
  \caption{a. shows a measurement of the phase of the AlGaAs coating at different chopper frequencies (blue pluses (+)). The red solid line shows a simulation of the phase of FS which is in good agreement with measurements on an uncoated part of the FS disc (light green crosses ($\times$)). The green dashed line shows the simulated phase for AlGaAs. In b. the corresponding amplitude measurement is shown (blue pluses). Using Equs.~\ref{equ:a1} this signal was split into the separate signals for FS (light blue dots ($\bullet$)) and AlGaAs (pink cosses). The red solid curve and the green dashed curve represent simulations for the amplitudes.}
  \label{fig:phase_FS_AlGaAs}
\end{figure}

Figure~\ref{fig:phase_FS_AlGaAs}\,a. shows the development of the phase with chopper frequency $p(f)$ for one position on the AlGaAs coating. The red solid line shows a simulation $p_{\rm FS}(f)$ for FS. The measurement of the phase of an uncoated part of the FS disc is shown by the light green crosses ($\times$) which is in agreement with the simulation. (The discrepancy for high frequencies is caused by low signal due to low/short heating). The green dashed line shows the theoretical frequency dependence of the phase for AlGaAs $p_{\rm{AlGaAs}}(f)$ and the dark blue pluses (+) represent the phase measurement for the crystalline AlGaAs coating on the FS disc which is a mixture of $p_{\rm{FS}}(f)$ and $p_{\rm{AlGaAs}}(f)$. Correspondingly the amplitude $a(f)$ will also be a mixture of the amplitudes of FS $a_{\rm{FS}}(f)$ and AlGaAs $a_{\rm{AlGaAs}}(f)$. The blue crosses ($\times$) in Fig.~\ref{fig:phase_FS_AlGaAs}\,b. show a measurement of the amplitude for the AlGaAs on FS system. The total (amplitude and phase) signal for this AlGaAs coating on FS system is given by

\begin{equation}
a(f) \begin{pmatrix} {\rm cos}(p(f)) \\ {\rm sin}(p(f)) \end{pmatrix} = a_{\rm{FS}}(f) \begin{pmatrix} {\rm cos}(p_{\rm{FS}}(f)) \\ {\rm sin}(p_{\rm{FS}}(f)) \end{pmatrix} + a_{\rm{AlGaAs}}(f) \begin{pmatrix} {\rm cos}(p_{\rm{AlGaAs}}(f)) \\ {\rm sin}(p_{\rm{AlGaAs}}(f)) \end{pmatrix}.
\label{equ:signal}
\end{equation}
From this we get the relations for the amplitudes

\begin{equation}
a_{\rm FS}(f) = \frac{a(f)({\rm sin}(p(f) - p_{\rm AlGaAs}(f)))}{{\rm sin}(p_{\rm FS}(f) - p_{\rm AlGaAs}(f))},
\enspace a_{\rm AlGaAs}(f) = \frac{a(f)({\rm sin}(p(f) - p_{\rm FS}(f)))}{{\rm sin}(p_{\rm AlGaAs}(f) - p_{\rm FS}(f))}.
\label{equ:a1}
\end{equation}
Using the phase information shown in Fig.~\ref{fig:phase_FS_AlGaAs}\,a. and Equs.~\ref{equ:a1} the measured amplitude $a$ for the mixed signal can be separated into the amplitude signals from the FS substrate $a_{\rm{FS}}$ and the AlGaAs coating $a_{\rm{AlGaAs}}$. The light blue dots in Fig.~\ref{fig:phase_FS_AlGaAs}\,b. represent the part of the signal originating from the FS disc and the pink crosses ($\times$) represent the part of the signal originating from the AlGaAs coating. The solid red and dashed dark green curves show corresponding simulations.

From a measured amplitude the absorption was calculated to be $\alpha_{\rm FS} = K \times a_{\rm FS}$. The calibration factor $K$ was obtained by using a FS calibration sample with known absorption. For the FS fraction of the amplitude signal shown in Fig.~\ref{fig:phase_FS_AlGaAs}\,b., the calibration which was made with the FS sample is valid. For the AlGaAs component of the amplitude signal the absorption has to be recalibrated taking the material parameters into account which are given in Tab.~\ref{tab:parameters}. An additional calibration factor calculates from

\begin{equation}
\frac{{\rm d}n/{\rm d}T_{\rm FS}}{{\rm d}n/{\rm d}T_{\rm AlGaAs}}\frac{k_{\rm AlGaAs}}{k_{\rm FS}} \times \frac{|DF_{\rm x,FS}|}{|DF_{\rm x,AlGaAs}|}
\end{equation}
to 1.04 at \unit[407]{Hz}, where DF$_{\rm x}$ is the dispersion function for surface absorption. For details see~\cite{Alexandrovski2009}. The measured absorption correspondingly calculates to

\begin{equation}
\alpha = K \times (\alpha_{\rm{FS}} + 1.04 \times \alpha_{\rm{AlGaAs}}).
\end{equation}

The calibration factor of 1.04 indicates that for the given material parameters and the measurement frequency, weighting of coating and substrate is very similar. This varies significantly for different materials and frequencies.

Figure~\ref{fig:phase_FS_AlGaAs}\,d. shows a map of the calibrated absorption $\alpha$ after recalibration for each single measurement depending on the varying phase. The mean value of the absorption is $\alpha_{\rm 1530} = \unit[(38.4 \pm 10.3)]{ppm}$ (mean value and standard deviation). In subsection~\ref{subsec:stack} this result is compared to the expected absorption for a highly reflective (HR) coating in which the vast majority of the light is reflected by only the first few layers of the coating.

\begin{table}
\centering
\small{
  \caption{Material parameters used for the simulations of amplitude and phase. The coating material parameters used for the simulation were the average of the values for GaAs and AlGaAs, weighted by the total thickness of each material in the coating stack.}
 \label{tab:parameters}
 \begin{tabular}{llll}
        \toprule
				Parameter																			&FS Substrate															&GaAs																		&AlGaAs\\
 				\midrule
 				index of refraction $n$	(\unit[1530]{nm})			&$1.44$~\cite{Leviton2006}								&$3.38$~\cite{Afromowitz}								&$2.93$~\cite{Afromowitz}\\
				index of refraction $n$	(\unit[1620]{nm})			&$1.44$~\cite{Leviton2006}								&$3.37$~\cite{Afromowitz}								&$2.93$~\cite{Afromowitz}\\
        thermo refr. coeff. d$n$/d$T$	[1/K]						&$8.57 \times 10^{-6}$~\cite{Weber2003}		&$2.04 \times 10^{-4}$~\cite{Kim2007}		&$1.74 \times 10^{-4}$~\cite{Kim2007}\\
        specific heat $c$ [J/(kg K)]									&$740$~\cite{Acuratus}										&$330$~\cite{Ioffe}											&$440.4$~\cite{Ioffe}\\
        density $\rho$ [$g/cm^3$]											&$2.34$~\cite{Weber2003}									&$5.32$~\cite{Ioffe}										&$3.98$~\cite{Ioffe}\\
        thermal expansion $a_{\rm th}$ [1/K]					&$0.55 \times 10^{-6}$~\cite{Weber2003}		&$5.73 \times 10^{-6}$~\cite{Ioffe}			&$5.24 \times 10^{-6}$~\cite{Ioffe}\\
        thermal conductivity $k_{\rm th}$ [W/(m\,K)]	&$1.38$~\cite{Weber2003, Acuratus}				&$55$~\cite{Ioffe}											&$70$~\cite{Ioffe}\\
				\bottomrule
 \end{tabular}
			}
 \end{table}

\subsection{Absorption of the (uncoated) fused silica disc}

\begin{figure}
  \centering
  \includegraphics[width=12cm]
    {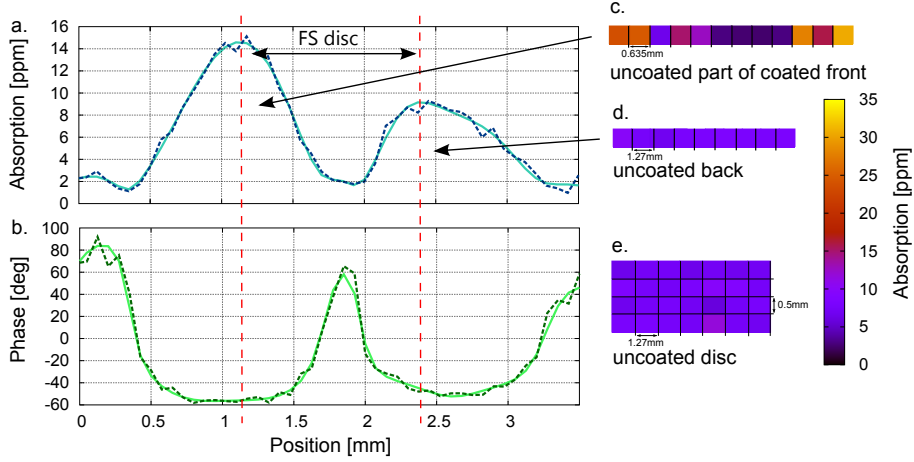}
  \caption{a. and b. show the PCI signal for calibrated amplitude (a.) and phase (b.) of an uncoated part of surface of the FS disc to which the AlGaAs coating is attached. The red (dashed) vertical line show the position of the disc's front and back surfaces. The front surface shows higher absorption than the back surface. c. shows several absorption measurements on the disc's front surface measured in a line with \unit[0.635]{mm} distance. Strong variations are visible with a mean value of $\unit[(13.1 \pm 10.7)]{ppm}$. d. shows an equivalent measurement of the absorption of the uncoated back surface. The absorption is more uniform with an average value of $\unit[(7.7 \pm 1.1)]{ppm}$. e. shows the absorption of an identical, but uncoated FS disc. The absorption of this surface is $\unit[(8.2 \pm 0.8)]{ppm}$ and in good agreement with the absorption of the back surface of the coated disc.}
  \label{fig:absorption_disc_messungen}
\end{figure}

In a further step the absorption of an uncoated part of the coated face of the FS disc was measured. Figure~\ref{fig:absorption_disc_messungen}a. shows the calibrated absorption signal of a scan through the FS disc, Fig.~\ref{fig:absorption_disc_messungen}b. shows the corresponding phase signal. The red dotted lines mark the position of the substrate's front (left) and back (right) surface. The absorption of the front surface is clearly higher than the absorption of the back surface. Therefore, we measured the absorption at several positions with \unit[0.625]{mm} separation on the front surface (Fig.~\ref{fig:absorption_disc_messungen}c.) and, after turning the disc around, repeated this measurement series for the back surface (Fig.~\ref{fig:absorption_disc_messungen}d.). These measurements confirmed that the absorption of the back surface is $\unit[(8.2 \pm 0.8)]{ppm}$ (mean value and standard deviation) and therefore significantly smaller and more uniform than on the front surface with $\unit[(13.1 \pm 10.7)]{ppm}$. For comparison we measured also the absorption of the surface of an identical, uncoated FS disc (Fig.~\ref{fig:absorption_disc_messungen}e.). The absorption of this surface was $\unit[(7.7 \pm 1.1)]{ppm}$ and in good agreement with the absorption of the back side of the coated substrate.

The origin of the absorption increase and whether we can assume a similar surface absorption increase under the coating is unknown. The only difference in storing and treatment of the samples front and back surface was the plasma cleaning procedure used prior to the bonding process. The surface roughness of the front surface was found to be higher than the back surface by \unit[25]{\%} which would not necessarily increase the absorption but might increase the susceptibility to contamination.

Therefore, we can only give an upper limit for the absorption of the coating stack by subtracting the absorption of $\unit[(8.2 \pm 0.8)]{ppm}$ of the back surface resulting in $\alpha_{\rm 1530} \leq \unit[(30.2 \pm 11.1)]{ppm}$. This corresponds to an absorption coefficient of $\alpha = \unit[0.06]{cm^{-1}}$ and an extinction coefficient of $k = 7.3 \times 10^{-7}$ at \unit[1530]{nm}.

\subsection{Intensity dependence of the absorption}

\begin{figure}
  \centering
  \includegraphics[width=10cm]
    {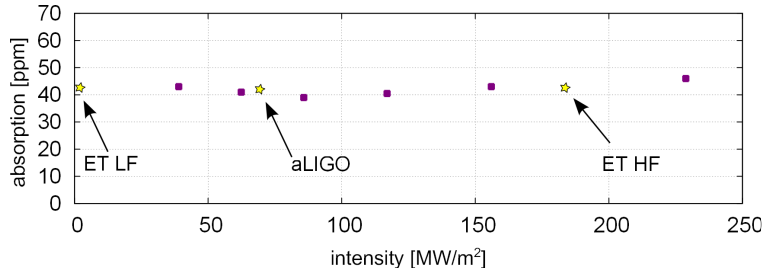}
  \caption{Optical absorption of one position of the AlGaAs coating for different laser light intensity: A beam radius of $\unit[35]{\mu m}$ was used for all measurements while the laser power was increased from \unit[150]{mW} to \unit[880]{mW}. The measured absorption is shown by the purple squares, the yellow stars show the design intensity within the arm cavities for ET and aLIGO.}
  \label{fig:intensity_dependence}
\end{figure}

Semiconductor materials such as AlGaAs show two-photon absorption effects in which two photons are absorbed simultaneously to enable a carrier to cross the band gap. The band gap for AlGaAs is between \unit[1.424]{eV} (\unit[871]{nm}, GaAs) and \unit[2.16]{eV} (\unit[574]{nm}, AlAs). Therefore, we measured the intensity dependence of the absorption at one position of the coating. The intensity was varied between $\unit[39]{MW/m^2}$ and $\unit[229]{MW/m^2}$. While the beam radius of $\unit[35]{\mu m}$ was constant for all measurements the laser power was increased from \unit[150]{mW} to \unit[880]{mW}. This intensity range is above the intensity for ET LF ($\unit[0.7]{MW/m^2}$) and includes the intensity for ET HF ($\unit[184]{MW/m^2}$) and aLIGO ($\unit[71]{MW/m^2}$). The measurement series is presented in Fig.~\ref{fig:intensity_dependence} and shows that the absorption of \unit[42]{ppm} varies by less than $\unit[10]{\%}$ within the studied intensity range. No significant increase at higher intensities was observed.

\subsection{Comparison to HR coating}
\label{subsec:stack}

The AlGaAs coating stack is optimized for high reflectivity at \unit[1064]{nm} while at \unit[1530]{nm} we measured a transmission of approximately \unit[70]{\%}.

The coating is composed of alternating layers of GaAs and AlGaAs each with an optical thickness of a quarter of the wavelength of $\lambda=\unit[1064]{nm}$. With refractive indices of $n_{\rm GaAs}=3.48041$ and $n_{\rm AlGaAs}=2.97717$ at \unit[1064]{nm} the optimal layer thickness would be \unit[76.4]{nm} for GaAs and \unit[89.3]{nm} for AlGaAs. The actual as-grown thicknesses of the layers are \unit[77.36]{nm} for GaAs and \unit[91.45]{nm} for AlGaAs.

At each layer boundary due to the change of refractive index light is reflected following Fresnel's formulas~\cite{Hecht2012}. To calculate the actual reflectivity and laser power within each double layer of the coating stack, the coating stack was simulated using ray-transfer matrix formalism for calculating a series of reflecting boundaries. Details can be found in~\cite{Steini2013}.

\begin{figure}
  \centering
  \includegraphics[width=9cm]
    {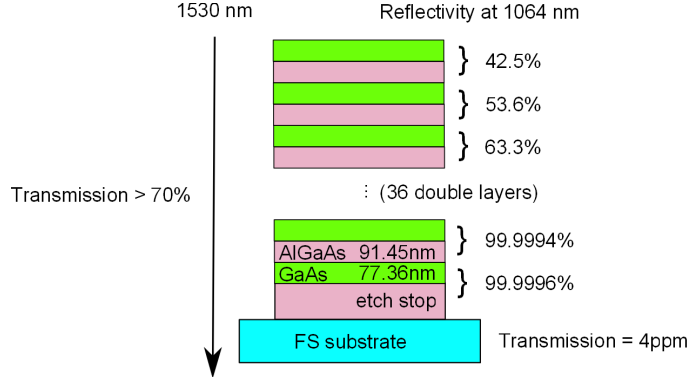}
  \caption{Schematic of the AlGaAs coating stack: The coating is bonded to a FS disc. The stack starts with a (thick) AlGaAs etch stop layer followed by 81 quarter wave layers of alternating GaAs/AlGaAs. The calculated reflectivity per double layer at \unit[1064]{nm} is given for the first three double layers showing that only a small part of the light propagates into the lower layers while at \unit[1530]{nm} a transmission of $\> \unit[70]{\%}$ was measured.}
  \label{fig:coating_stack}
\end{figure}

The resulting reflectance of the coating stack per double layer shows that at \unit[1064]{nm} \unit[42.5]{\%} of the input power is already reflected at the first double layer, \unit[53.6]{\%} at the second double layer etc. as indicated in Fig.~\ref{fig:coating_stack}. So the power within the first double layer is reduced to \unit[(100 - 42.5)]{\%}. The sum of the power per double layer results in an effective penetration depth of $D_{\rm eff} = (1-0.425)+(1-0.536)+(1-0.633)+... =2.6$ double layers of the input laser power into the coating stack. The total absorption measured would thus be equivalent to the absorption from 2.6 double layers which saw the full power of the laser. Repeating this calculation for a wavelength of \unit[1530]{nm} results in $D_{\rm eff 1530} = 29.6$ double layers.

The ratio $D_{\rm eff 1530}/D_{\rm eff 1064}$ gives the scaling which is required to make the measured absorption comparable to the expected absorption for a stack designed to be highly reflective at \unit[1530]{nm} with the additional consideration that the quarter wave single layers will be thicker by $\unit[1530]{nm}/\unit[1064]{nm}=1.4$. Scaling the measurement using these factors results in $\alpha_{\rm HR} = 1.4/11.6 \times \alpha_{\rm 1530} \leq \unit[(3.6 \pm 1.3)]{ppm}$ which is in good agreement with previous results at \unit[1064]{nm} and \unit[1550]{nm}~\cite{Cole2014}.

\section{Summary and Conclusion}

We have used photo-thermal common-path interferometry to measure the optical absorption at \unit[1530]{nm} of an AlGaAs crystalline coating stack, designed for high reflectivity at \unit[1064]{nm} and bonded to a FS disk.

The AlGaAs coating bonded to FS showed an unexpectedly large signal originating directly from the coating, requiring the simulation and analysis of the signal to be adapted.

To check the uniformity of the coating, an area of \unit[4]{mm} $\times$ \unit[5.7]{mm} was mapped. The absorption of the complete stack as would be seen by the laser beam in a GW detector when transmitting the ITM was measured to be $\alpha_{\rm 1530} = \unit[(38.4 \pm 10.3)]{ppm}$. This includes substrate surface absorption and a possible absorption of the bond in between substrate and coating. Subtracting the lower limit for the substrate surface absorption, $\alpha_{\rm 1530} \leq \unit[(30.2 \pm 11.1)]{ppm}$ forms an upper limit for the coating absorption. The corresponding absorption coefficient is $\alpha = \unit[0.06]{cm^{-1}}$ ($k = 7.3 \times 10^{-7}$). This is equivalent to an absorption of $\alpha_{\rm HR} \leq \unit[(3.6 \pm 1.3)]{ppm}$ in an AlGaAs stack designed for HR at \unit[1530]{nm}. This is a very promising low absorption result for alternative low--loss coating materials. Still, further work will be necessary to reach the requirements of $< \unit[1]{ppm}$ for future gravitational wave detectors.

Measurements of the phase show evidence of non-uniformity in the thermal contact between the coating and the substrate, which may indicate variations in the quality of the bond. These may be related to a insufficient quality of the substrate polish.
The surface of the fused silica disk to which the coating was attached showed an increased absorption compared to the back surface of the sample. This is unexplained, but may be related to the plasma cleaning procedure used prior to bonding the coating.

One possible source of the measured AlGaAs absorption could be contaminants from the MBE growth process, which was carried out in a chamber used to deposit various materials including doped semiconductor layers. A dedicated AlGaAs MBE system, with reduced potential for contamination, may enable higher purity, and thus possibly lower absorption, coatings to be produced.

\section*{Acknowledgements}

This work was partly funded by a Royal Society Research Grant (RG110331), and we are grateful for additional financial support from STFC and the University of Glasgow. IWM is supported by a Royal Society Research Fellowship. SR holds a Royal Society Wolfson Research Merit award. We are grateful to the International Max Planck Partnership for Measurement and Observation at the Quantum Limit for support, and we thank our colleagues in the LSC and VIRGO collaborations and within SUPA for their interest in this work. We thank Alexei Alexandrovski for valuable discussions.

This paper has LIGO Document number LIGO-P1400226.
%
%

\end{document}